\documentclass[conference]{IEEEtran}
\IEEEoverridecommandlockouts
\usepackage{cite}
\usepackage{amsmath,amssymb,amsfonts}
\usepackage{algorithmic}
\usepackage{graphicx}
\usepackage{textcomp}
\usepackage{xcolor}
\usepackage{url}
\usepackage[hidelinks]{hyperref}
\usepackage{float}
\usepackage{subcaption}
\usepackage{array}
\def\BibTeX{{\rm B\kern-.05em{\sc i\kern-.025em b}\kern-.08em
    T\kern-.1667em\lower.7ex\hbox{E}\kern-.125emX}}
\begin{document}

\title{Memory-Disaggregated In-Memory Object Store Framework for Big Data Applications}

\author{\IEEEauthorblockN{Robin Abrahamse}
\IEEEauthorblockA{\textit{Accelerated Big Data Systems} \\
\textit{Delft University of Technology}\\
Delft, The Netherlands \\
\href{mailto:R.Abrahamse@student.tudelft.nl}{R.Abrahamse@student.tudelft.nl}}
\and
\IEEEauthorblockN{\'Akos Hadnagy}
\IEEEauthorblockA{\textit{Accelerated Big Data Systems} \\
\textit{Delft University of Technology}\\
Delft, The Netherlands \\
\href{mailto:A.Hadnagy@tudelft.nl}{A.Hadnagy@tudelft.nl}}
\and
\IEEEauthorblockN{Zaid Al-Ars}
\IEEEauthorblockA{\textit{Accelerated Big Data Systems} \\
\textit{Delft University of Technology}\\
Delft, The Netherlands \\
\href{mailto:Z.Al-Ars@tudelft.nl}{Z.Al-Ars@tudelft.nl}}
}

\maketitle

\begin{abstract}
The concept of memory disaggregation has recently been gaining traction in research. With memory disaggregation, data center compute nodes can directly access memory on adjacent nodes and are therefore able to overcome local memory restrictions, introducing a new data management paradigm for distributed computing. This paper proposes and demonstrates a memory disaggregated in-memory object store framework for big data applications by leveraging the newly introduced ThymesisFlow memory disaggregation system. The framework extends the functionality of the pre-existing Apache Arrow Plasma object store framework to distributed systems by enabling clients to easily and efficiently produce and consume data objects across multiple compute nodes. This allows big data applications to increasingly leverage parallel processing at reduced development costs. In addition, the paper includes latency and throughput measurements that indicate only a modest performance penalty is incurred for remote disaggregated memory access as opposed to local ($\sim$6.5 vs $\sim$5.75 GiB/s). The results can be used to guide the design of future systems that leverage memory disaggregation as well as the newly presented framework. This work is open-source and publicly accessible at \url{https://doi.org/10.5281/zenodo.6368998}.\end{abstract}

\section{Introduction}
\begin{figure}[H]
    \begin{subfigure}[t]{0.23\textwidth}
        \centering
        \includegraphics[width=\linewidth]{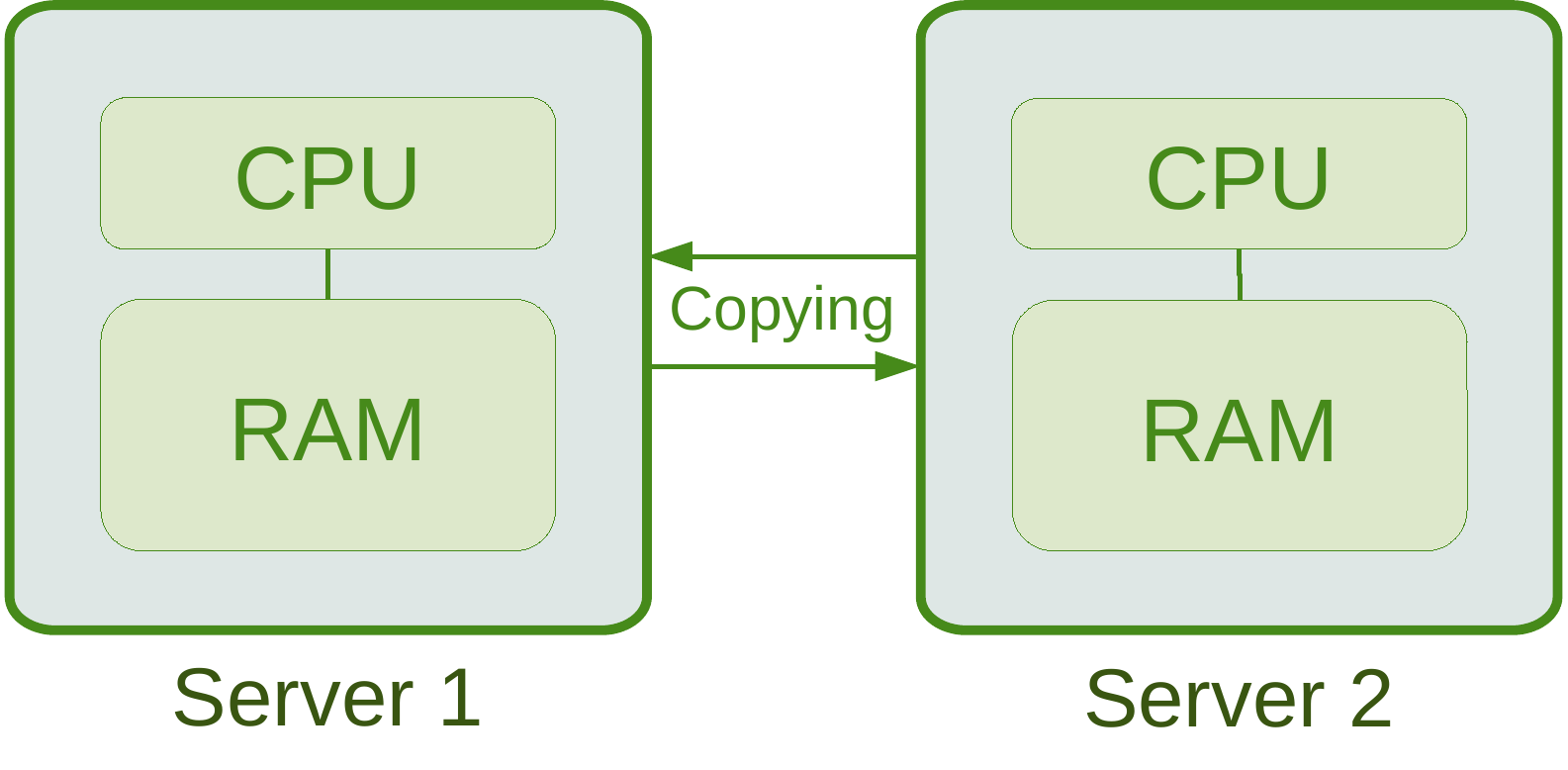}
        \caption{Scale-out}
        \label{fig:scale-out}
    \end{subfigure}
    \hfill
    \begin{subfigure}[t]{0.23\textwidth}
        \centering
        \includegraphics[width=\linewidth]{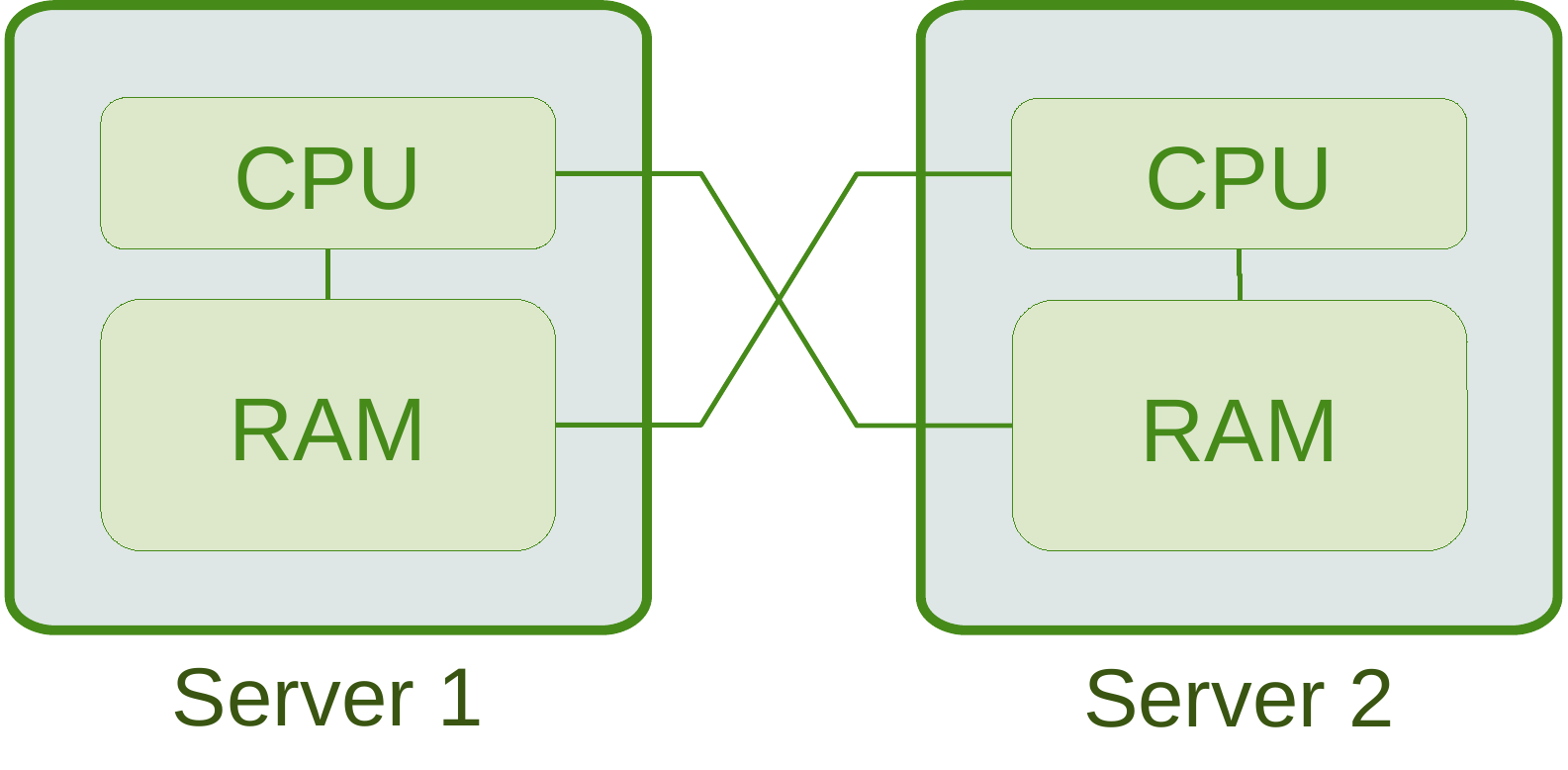}
        \caption{Memory Disaggregated}
        \label{fig:mem-disagg}
    \end{subfigure}
    \caption{Distributed Computation Scaling Approaches}
\end{figure}

Big data workloads are often limited in scale by the memory volume available to local systems. Expanding memory volume in data center servers is associated with super-linear costs~\cite{memdisag_problems} and consequently, a scale-out approach is commonly used to scale data center applications for larger data volumes. In a scale-out approach, vast amounts of data are sent over the local network and copied to local memory (Figure~\ref{fig:scale-out}), contending for network bandwidth and often harming performance by thrashing memory across the compute nodes~\cite{memdisag_problems}. 

In this paper, we propose a framework that leverages memory disaggregation technology to mitigate these issues by providing direct access to large memory volumes over a custom network (Figure~\ref{fig:mem-disagg}), relaxing the need to utilize scarce local network bandwidth and the burden to evict data from memory in order to copy. This framework also allows distributed systems to increase their ability to parallelize data processing.

Memory disaggregation refers to the decoupling of directly accessible memory from individual compute nodes (e.g. servers in a data center rack). Specifically, it entails the integration of hardware-enabled systems which allow compute nodes to directly access memory from other -- remote -- compute nodes and therefore increase effective memory volume (Figure~\ref{fig:mem-disagg}). 

By enabling compute nodes to access and modify both local and remote disaggregated memory concurrently, without duplicating data, memory disaggregation has the potential to improve application performance. Compute nodes could for example operate on local in-memory data while utilizing in-memory data from the other nodes in the network (i.e. wide-dependency operations). This increases the ability of data center applications to process data in a parallel manner.

The current work proposes a memory disaggregated framework for big data applications and evaluates its performance. This paper contains the following main contributions:
\begin{itemize}
    \item The proposal and implementation of a memory disaggregated in-memory object store framework prototype, which enables easy and efficient production and consumption of data objects across distributed compute nodes.
    \item A set of microbenchmarks and respective results for measuring latency and throughput of creating and retrieving data objects from the proposed object store framework.
    \item Considerations and recommendations for future work on the newly proposed framework and memory disaggregation technology in general.
\end{itemize}
The paper researches the performance-enhancing potential and application of memory disaggregation for big data analytics with ThymesisFlow~\cite{thymesis} and provides a stepping stone for future work.

This paper is organized as follows: first, Section~\ref{relatedwork} outlines related work. Section~\ref{background} provides additional background information on the used memory disaggregation prototype and workloads. Then, Section~\ref{methodology} presents the proposed memory disaggregated object store framework and corresponding set of microbenchmarks. Lastly, Section~\ref{results} discusses the results and future work before Section~\ref{conclusion} concludes the paper.

\section{State of the Art and Related Work}\label{relatedwork}
\subsection{ThymesisFlow}
At the time of writing, the field of memory disaggregated big data analytics is still in an exploratory phase but is already gaining traction in research~\cite{memdisag_problems,memdisag_sup1,memdisag_sup2}. Pre-existing technologies have thus far only demonstrated varying results on smaller systems. However, recent work by IBM introduced a promising memory disaggregation framework for data center infrastructures, called \emph{ThymesisFlow}~\cite{thymesis}. ThymesisFlow allows servers to transparently access memory from adjacent servers through a custom network, bypassing local memory volume restrictions.

ThymesisFlow offers an initial step towards scalable memory disaggregation for big data applications~\cite{webinar}. It presents an early opportunity to test the potential of memory disaggregation for acceleration of data center workloads. This could guide the design of future server-scale and rack-scale hardware systems as well as novel programming models and applications. 

Through efficient pooling of processing and memory resources, this technology has the potential to increase data center utilization rates, decrease total cost of data center ownership, and improve both performance and cost of development for applications~\cite{thymesis}. Consequently, it can contribute to improving the efficiency of data center workloads generally and big data workloads in particular. Additionally, due to the simple memory sharing interface, workloads could be scaled and parallelized more easily.

In fact, similar technology is already being incorporated in the next generation of data center processors. IBM has announced an improved memory disaggregation function to be integrated into upcoming IBM POWER10 processors under the name 'Memory Inception'~\cite{power10}.

\subsection{Plasma Object Store}
Big data applications often consume data from external sources and acquire this data by querying the source. A single source may have multiple consumers querying it. The Apache Arrow framework~\cite{arrow} aims to standardize this consumer-supplier dynamic in an efficient way by providing tools to share in-memory data between applications without serialization overhead. Part of this framework is the Plasma in-memory object store, which is used to store and access immutable data objects within a system in which multiple data suppliers and consumers may exist.

The Plasma object store lives as a separate process to which clients of the store may commit and 'seal' data objects with an object identifier. The store manages the objects' locations in shared memory and makes them available to other clients upon sealing. Sealing an object prompts the store to make it immutable, such that race conditions cannot occur. Big data applications often do not require mutability of the source data and can therefore benefit from the reduced concurrency complexity, e.g. the Resilient Distributed Dataset (RDD) of Apache Spark~\cite{spark} is also built on this premise.

Plasma store clients can access the existing sealed objects by querying the store for the object identifiers. The store then provides the client with a read-only buffer containing the object data, which the client can consecutively consume. Sharing through system memory ensures that both object commitment and access incur only marginal latency penalties. Moreover, the standardized format of the store eliminates serialization overhead between processes which improves performance and efficiency. The framework is already being leveraged in existing big data workloads~\cite{arrowsam}.

\section{Technological Background}
\label{background}
ThymesisFlow was developed for POWER9~\cite{power9} architectures and leverages the OpenCAPI~\cite{opencapi} interface. The system builds on FPGA accelerators that interface between the ThymesisFlow network and the Linux operating system kernel. The disaggregated memory is exposed to the operating system as a memory region, which is accessible through the ThymesisFlow system such that it becomes transparent to applications~\cite{thymesis}.

\begin{figure}[ht]
    \centering
    \includegraphics[width=.6\linewidth]{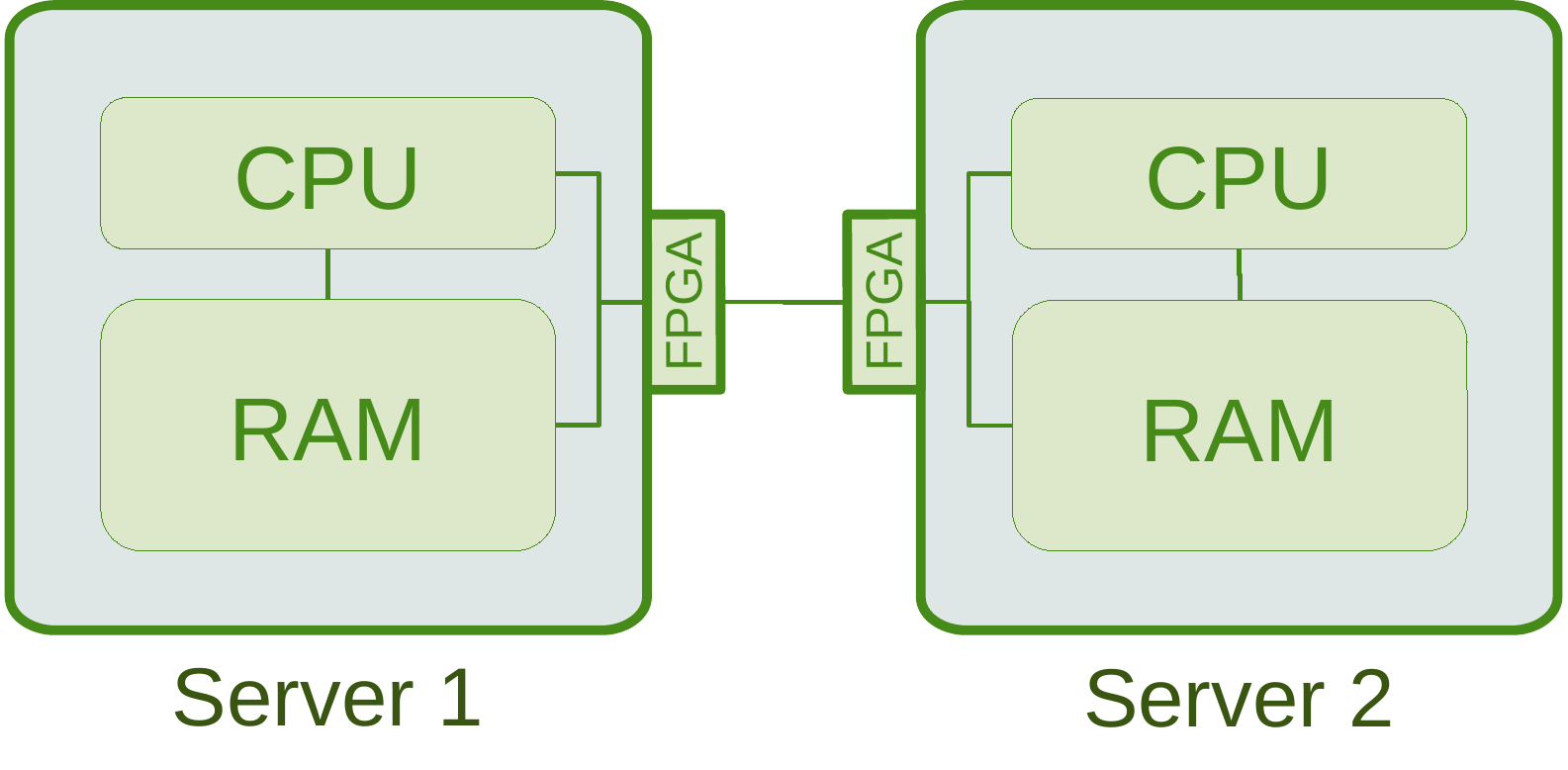}
    \caption{ThymesisFlow Schematic Representation}
    \label{fig:thymesis}
\end{figure}

Figure~\ref{fig:thymesis} shows a schematic representation of the effective physical flow of data in ThymesisFlow. Memory sharing happens through the FPGA accelerators, leveraging the OpenCAPI FPGA stack. OpenCAPI~\cite{opencapi}, is an interface architecture that enables accelerators to cache-coherently access system memory. This means that the FPGA accelerator can access memory from the host in a cache-coherent way.

For ThymesisFlow, this entails that a portion of local system memory is marked as disaggregated and made available to remote compute nodes. The system's FPGA accelerator then mimics a memory controller of sorts for remote disaggregated memory regions. Load and store instructions targeting the disaggregated memory region are relayed to the FPGA, which translates memory addresses and requests the appropriate memory regions from remote compute nodes. The remote compute node's FPGA then uses the OpenCAPI interface to retrieve cache-coherent data from the desired memory regions and returns this data to the requesting node. This completes the call to remote disaggregated memory.

As it stands currently, memory disaggregation with ThymesisFlow is subject to several drawbacks as well. For example, calls to disaggregated memory carry an inherent latency penalty due to the extra distance data has to travel off-chip relative to local data. This penalty has been observed to be non-negligible~\cite{thymesis}, thus local memory remains of importance for performant applications~\cite{memdisag_problems}. This extra distance has to be traversed in scale-out approaches as well and, with efficient custom hardware and network protocols, memory disaggregation could even reduce this latency penalty with respect to traditional local networking.

Moreover, compute nodes using ThymesisFlow could potentially suffer from increased cache-coherency and synchronization issues related to the distributed nature of disaggregated memory fetching. When multiple processes are accessing and modifying the same memory region, cached changes need to be accessible to all processes. Usually, cache-coherency among shared-memory processes is handled by the operating system kernel, however, with memory disaggregation multiple operating systems are involved. Cache-coherency in this setting is not supported by common operating systems and eliminating caching completely -- which would require the development of custom kernel modules -- comes at a cost. Additionally, the increased latency of disaggregated memory calls begs extra caution with race conditions.

\begin{figure}[H]
    \begin{subfigure}[t]{0.23\textwidth}
        \centering
        \includegraphics[width=\linewidth]{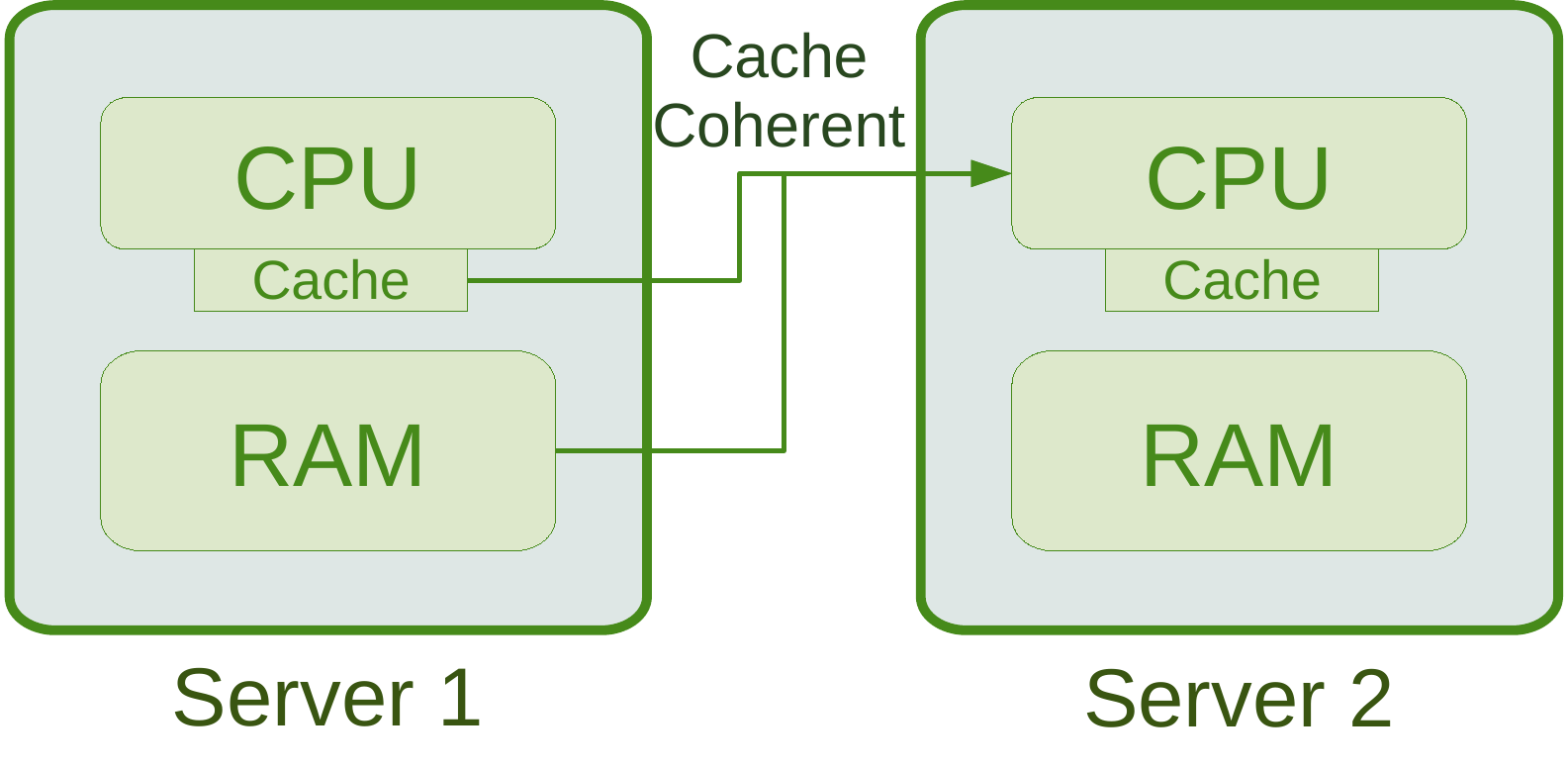}
        \caption{Remote Read}
        \label{fig:cc}
    \end{subfigure}
    \hfill
    \begin{subfigure}[t]{0.23\textwidth}
        \centering
        \includegraphics[width=\linewidth]{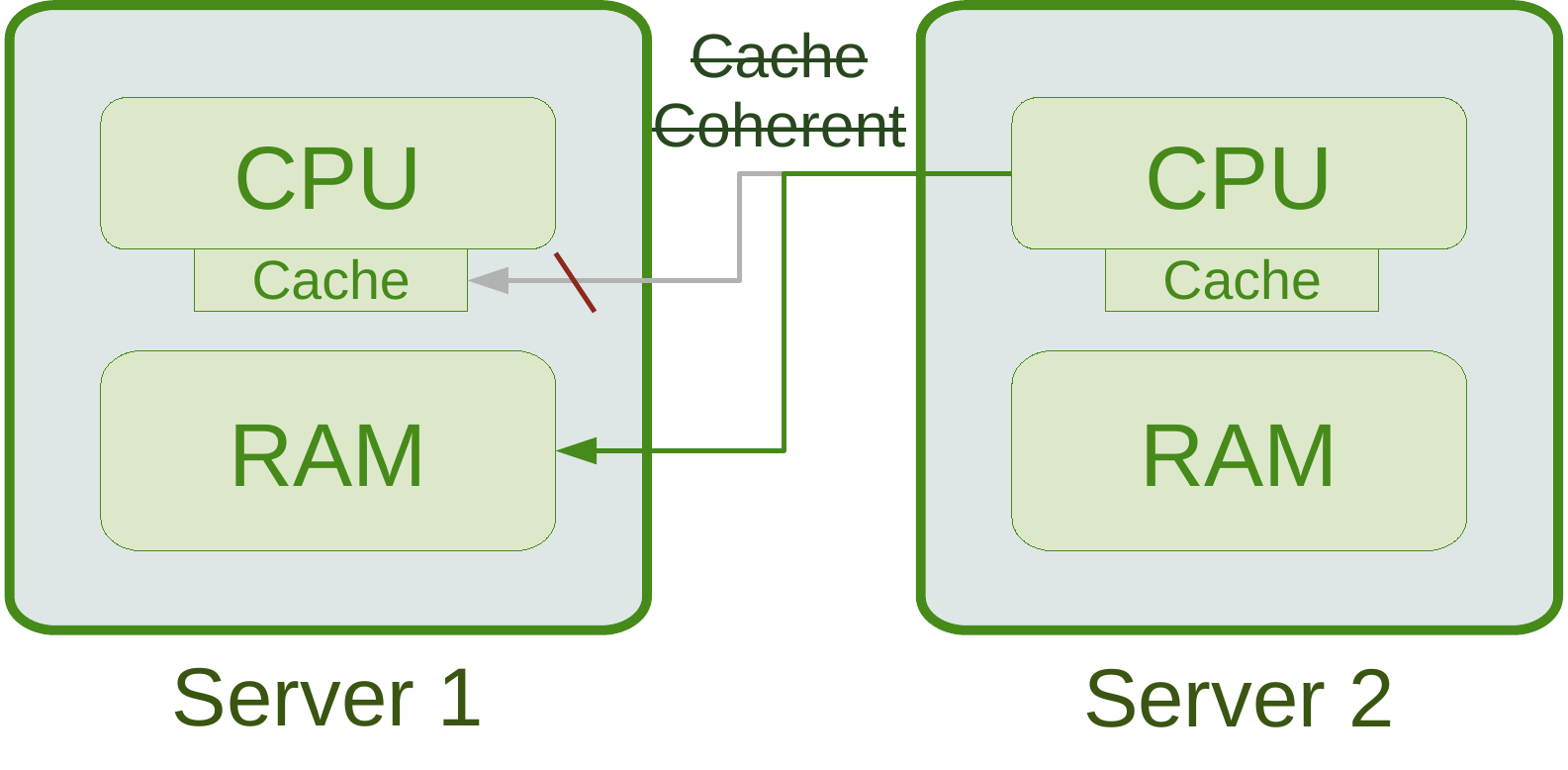}
        \caption{Remote Write}
        \label{fig:ncc}
    \end{subfigure}
    \caption{Cache-Coherency in ThymesisFlow Transactions}
    \label{fig:cache}
\end{figure}

The cache-coherency concerns arise from the data flow within ThymesisFlow and have important implications for its usage. The OpenCAPI~\cite{opencapi} interface ensures that reading remote disaggregated memory is cache-coherent (Figure~\ref{fig:cc}). Alternatively, writing to remote disaggregated memory is cache-coherent with the local system, but not necessarily with the remote system to which is being written. Explicitly, this means that data written to remote disaggregated memory is not necessarily immediately available to applications on the remote system. The written data will be flushed to the remote disaggregated memory, however, the remote system may have cached a previous value (Figure~\ref{fig:ncc}). This has implications for the memory disaggregated Plasma store, which will be discussed in the next section. Note that Figure~\ref{fig:cache} is only a conceptual schematic representation.

\section{Memory Disaggregated Object Store} \label{methodology}
The Apache Arrow Plasma object store~\cite{arrow} is currently restricted by the fact that it only supports local object storage. This means that the availability of both memory volume and processing units is limited. Memory disaggregation provides an opportunity to enable Plasma to access significantly larger volumes of memory and increase the number of available processing units.

The goal of the current work is to propose and evaluate a variant of the Apache Arrow Plasma object store~\cite{arrow} that leverages memory disaggregation through ThymesisFlow~\cite{thymesis}. The object store uses a strategy similar to sharding to leverage disaggregated memory, where objects are distributed over several servers and clients can directly retrieve objects from their memories. The development was conducted in two distinct stages: 
\begin{enumerate}
    \item Integration of ThymesisFlow into the original Plasma framework.
    \item Benchmarking of the novel Plasma framework.
\end{enumerate}
This paper predominantly focuses on the first stage to provide overall architecture for the proposed framework. The details of the two stages will be elaborated on further in this section. The source code can be found at \url{https://doi.org/10.5281/zenodo.6368998}.

\subsection{ThymesisFlow Plasma Store Integration}
The integration of ThymesisFlow in Plasma can be achieved in two separate steps:
\begin{enumerate}
    \item Disaggregated memory allocation; allows the Plasma store to allocate objects in local disaggregated memory such that they can be accessed by remote clients without resorting to slow scale-out approaches or network strain.
    \item Remote object sharing; sharing the objects contained in Plasma stores allows them to be retrieved by clients on all compute nodes.
\end{enumerate}

\subsubsection{Disaggregated Memory Allocation}
As an initial step, the Plasma store was modified to allocate objects in local disaggregated memory. Since the Plasma store is essentially a memory bookkeeping service for Plasma data objects, it requires sufficiently performant memory allocation. Originally, Plasma uses the Doug Lea Malloc library (dlmalloc)~\cite{dlmalloc} for this purpose, together with a file descriptor system to coordinate memory-mapping across Plasma store and clients. This ensures portability, among other desirable aspects, but does not suit the purpose of allocating in disaggregated memory.

Since ThymesisFlow is inherently Linux-based, the loss of portability by substitution of dlmalloc is not a limiting factor. Hence, dlmalloc was replaced by a simple allocation algorithm that receives the memory-mapped local disaggregated memory region and uses it to allocate Plasma objects. The algorithm simply allocates a chunk of memory to the first available region that can accommodate it. By using an ordered map data structure with logarithmic time look-up to keep track of the sizes of available regions, performance should not suffer critically. The replacement allocator does not consider e.g. locality, alignment, and fragmentation in memory allocation and thus surrenders some benefits to the original dlmalloc library~\cite{dlmalloc}. It should, however, suffice for exploring the performance of memory disaggregated systems.

\subsubsection{Remote Object Sharing}
Plasma conducts Inter-Process Communication (IPC) between Plasma store and clients through Unix domain sockets. This means that Plasma clients cannot directly communicate with remote Plasma stores as the latter exist in different operating systems, unreachable by the Unix domain sockets. An additional infrastructure could be created to accommodate for this, however, that would require all clients to connect with remote Plasma stores and would cause large amounts of duplicate data to be sent over the network. Therefore, it is more efficient to interconnect Plasma stores. Additionally, this means that the distributed nature can largely remain hidden to Plasma clients.

Sharing remote Plasma objects between stores introduces several additional constraints to the system. Obviously, Plasma stores must be able to communicate with each other about currently existing objects. Two significant new constraints were identified:
\begin{itemize}
    \item Identifier uniqueness; object identifiers must be unique across the system of all connected Plasma stores to prevent ambiguity.
    \item Distributed object-usage sharing; Plasma stores should have up-to-date information about which of their local objects are in use by clients system-wide.
\end{itemize}

The requirement for identifier uniqueness is an immediate consequence of the distributed nature of the proposed framework. If object identifiers are not unique across all connected Plasma stores, then this may introduce ambiguity in client requests for objects such that clients will not be able to retrieve all objects available in the Plasma stores.

The distributed object-usage sharing constraint relates to a Plasma store's internal policy about evicting objects when needed. Locally, the Plasma store keeps track of which objects are in use by its connected clients. In-use objects will not be evicted, because clients might still be reading from memory and evicting the objects would likely corrupt their data. Object-usage tracking should be extended to adopt this functionality across multiple Plasma stores. In the scope of the current work, this constraint was considered, however, was not limiting for demonstrative purposes and no solution was implemented yet.

A rough subdivision can be made in the possible approaches used to share Plasma object information across the system:
\begin{itemize}
    \item A shared data structure in disaggregated memory; Plasma stores can share their data structure that maps object identifiers to their corresponding buffers.
    \item Messaging via disaggregated memory; Plasma stores can perform point-to-point messaging between each other through disaggregated memory.
    \item Sharing via LAN; Plasma stores can communicate over the local network using common networking techniques.
\end{itemize}
The type of approach has far-reaching implications for the system architecture. These implications will be discussed next.

The first approach would allow remote Plasma stores to efficiently look up whether an object already exists and find its corresponding buffer. This approach requires handling issues specific to the usage of shared memory data structures such as preventing (local) heap allocation. Moreover, since this is a one-way (local Plasma store polls remote Plasma store) communication system, there is no efficient way for the local store to feed back information on currently in-use objects for the eviction policy. The latter relates to the previously discussed fact that writing to remote disaggregated memory may lead to cache-coherency issues on the remote compute node and doing so regardless would open the door to unfavorable race conditions. This could be accommodated by designing a kernel module to disable the memory caching behavior, but that is outside the scope of the current work.

For the second approach, a messaging system could be implemented, similar to the system suggested in~\cite{webinar}. Messaging in traditional shared memory is a simple task, however, the cache-coherency characteristics of ThymesisFlow introduce additional complexity. This would require developing a robust messaging system using both local and remote disaggregated memory. Any potential performance gain relative to using existing LAN techniques would be marginal considering communication protocol overhead and would incur significant additional development costs. A hybrid system that combines disaggregated memory hash map look-up with messaging could yield more favorable results, but this is also outside the scope of the current work.

\begin{figure}[H]
    \centering
    \includegraphics[width=.7\linewidth]{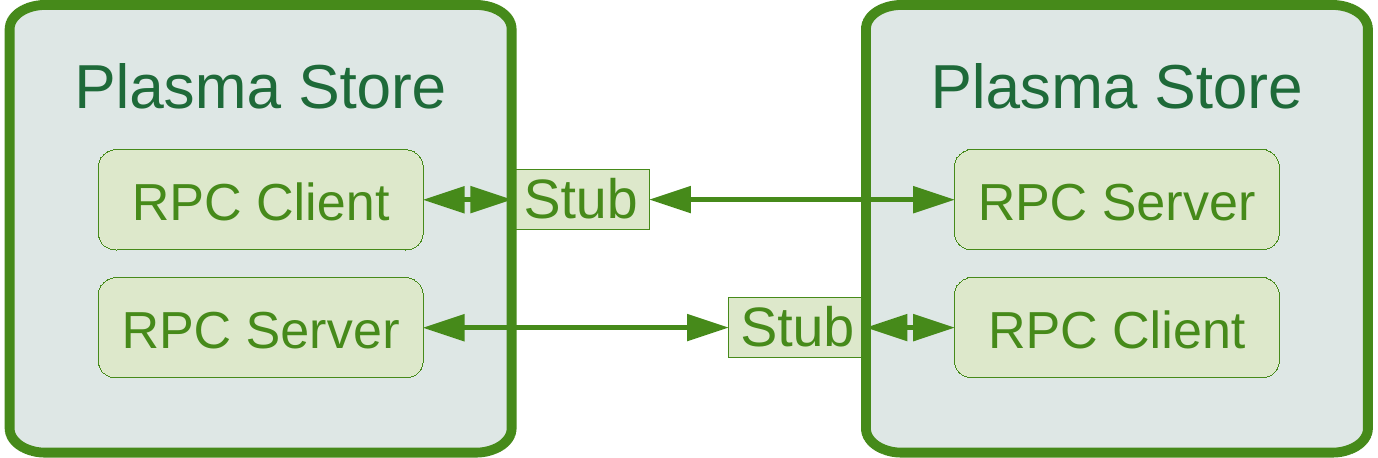}
    \caption{gRPC Functionality}
    \label{fig:grpc}
\end{figure}

Lastly, the third approach could be implemented in several ways as well. A simple, robust, and performant approach to do this is based on the Remote Procedure Call (RPC) concept. In this concept, an application can call a function through an RPC client as if it was executed by a remote application, hiding the networking complexity from the application. An efficient implementation for HPC is gRPC~\cite{grpc}, which is a high-performance RPC based on Protocol Buffers and HTTP/2~\cite{grpc_perf}. Internally, a gRPC client connects a stub to a remote gRPC server and relays the local function call to the server, which executes the call and returns the result~\cite{grpc}. This RPC functionality could be used to satisfy the constraints outlined before by allowing servers to look up objects remotely (Figure~\ref{fig:grpc}).

Considering the previously discussed drawbacks of disaggregated memory communication and the simplicity and performance of LAN communication, gRPC (version 1.38.0) was used to share objects between Plasma stores. This has implications for the remainder of the system.

The gRPC protocol was configured in synchronous mode due to its favorable servicing latency. As a result, the gRPC server requires a dedicated thread to service all calls synchronously. Additionally, gRPC was configured in unary mode to minimize protocol overhead for the messages being sent around.

The gRPC functionality works as follows: upon a client request for a remote object, the local Plasma store makes an RPC call to look up the object identifier(s) in the remote store, which contains a map data structure for object identifiers. Consequently, the local Plasma store retrieves the corresponding object buffer(s) and passes them on to the client. Similarly, on object creation, RPC calls are used to ensure the uniqueness of object identifiers.

This multithreaded look-up introduces the need for thread-safety mechanisms as both the Plasma store main thread and gRPC server thread may attempt to access the local object identifier map concurrently. Mutex functionality was built in to ensure thread-safety and eliminate race conditions that might otherwise occur.

\begin{figure}[ht]
    \centering
    \includegraphics[width=\linewidth]{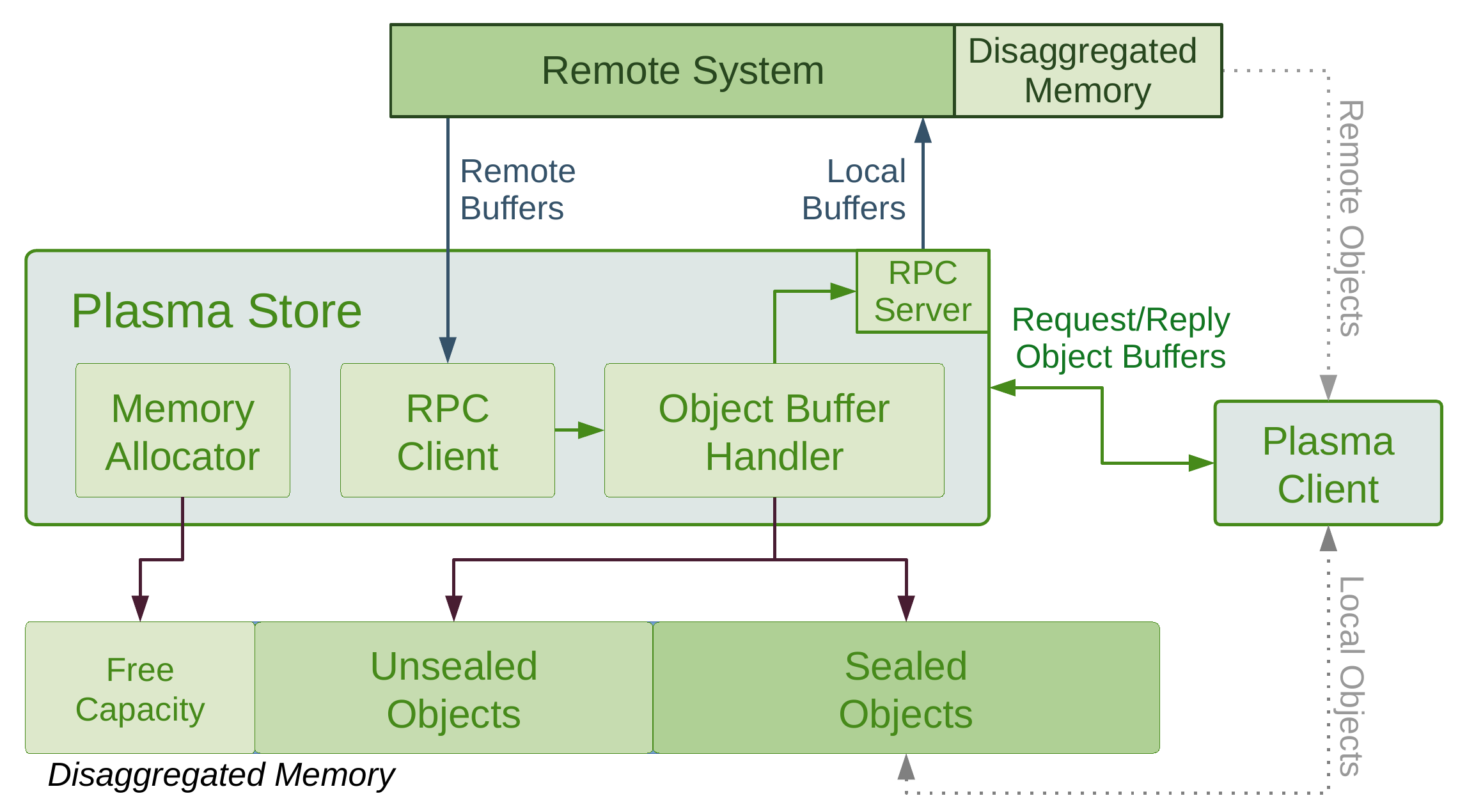}
    \caption{Memory Disaggregated Plasma Framework}
    \label{fig:blockdiagram}
\end{figure}

A schematic block diagram with full the proposed system can be found in Figure~\ref{fig:blockdiagram}.

\subsection{Benchmarking}\label{benchmarks}
The current work includes a set of microbenchmarks to quantify the difference in latency and throughput between local and remote Plasma object consumption. The benchmarks are designed to test a system of at least 2 Plasma stores. 

The benchmarks commit Plasma objects with random data to one of the Plasma stores, after which both local and remote clients will then request these objects' buffers from their local Plasma stores. The clients then receive the corresponding object buffers from the local Plasma store and retrieve their data sequentially. The data contents of the objects should not influence the system performance. 

In this process, several distinct measurements are done. Firstly, creation, writing, and sealing of the objects is measured. The consequent retrieval of object buffers by the client and reading of these object buffers are then measured separately. The benchmarks provide information about system performance and aim to guide the design of future systems leveraging memory disaggregation and Plasma.

6 different microbenchmarks were included to investigate performance in different scenarios, each repeated 100 times to monitor the effect of jitter in the system. The benchmarks test the Plasma framework with different orders of magnitude in object sizes and also vary the number of objects created. This way, they can capture differences in local and remote memory performance and variability of full-system performance with different object sizes. The number of objects is varied to mitigate any potential influence of caching of smaller objects. The specifications for each benchmark can be found in Table~\ref{tab:benchmarks}.

\begin{table}[H]
\centering
\setlength\extrarowheight{2pt}
\begin{tabular}{ c  p{0.15\linewidth}  p{0.16\linewidth} }
    \hline
     & \textbf{Number of Objects} & \textbf{Object Size (kB)}\\
    \hline
    \textit{1} & 1000 & 1 \\ 
    \textit{2} & 500 & 10 \\  
    \textit{3} & 200 & 100 \\
    \textit{4} & 100 & 1000 \\
    \textit{5} & 50 & 10000 \\
    \textit{6} & 10 & 100000 \\
    \hline
\end{tabular}
\caption{Benchmark Specifications}
\label{tab:benchmarks}
\end{table}

\section{Results \& Discussion}\label{results}
\subsection{Benchmarks}
The microbenchmarking experiment was run on two IBM Power System IC922 in combination with Alpha Data ADM-PCIE-9V3 FPGAs. All the benchmarks described in Section~\ref{benchmarks} were run on a single thread.

\begin{figure}[H]
    \centering
    \includegraphics[width=1\linewidth]{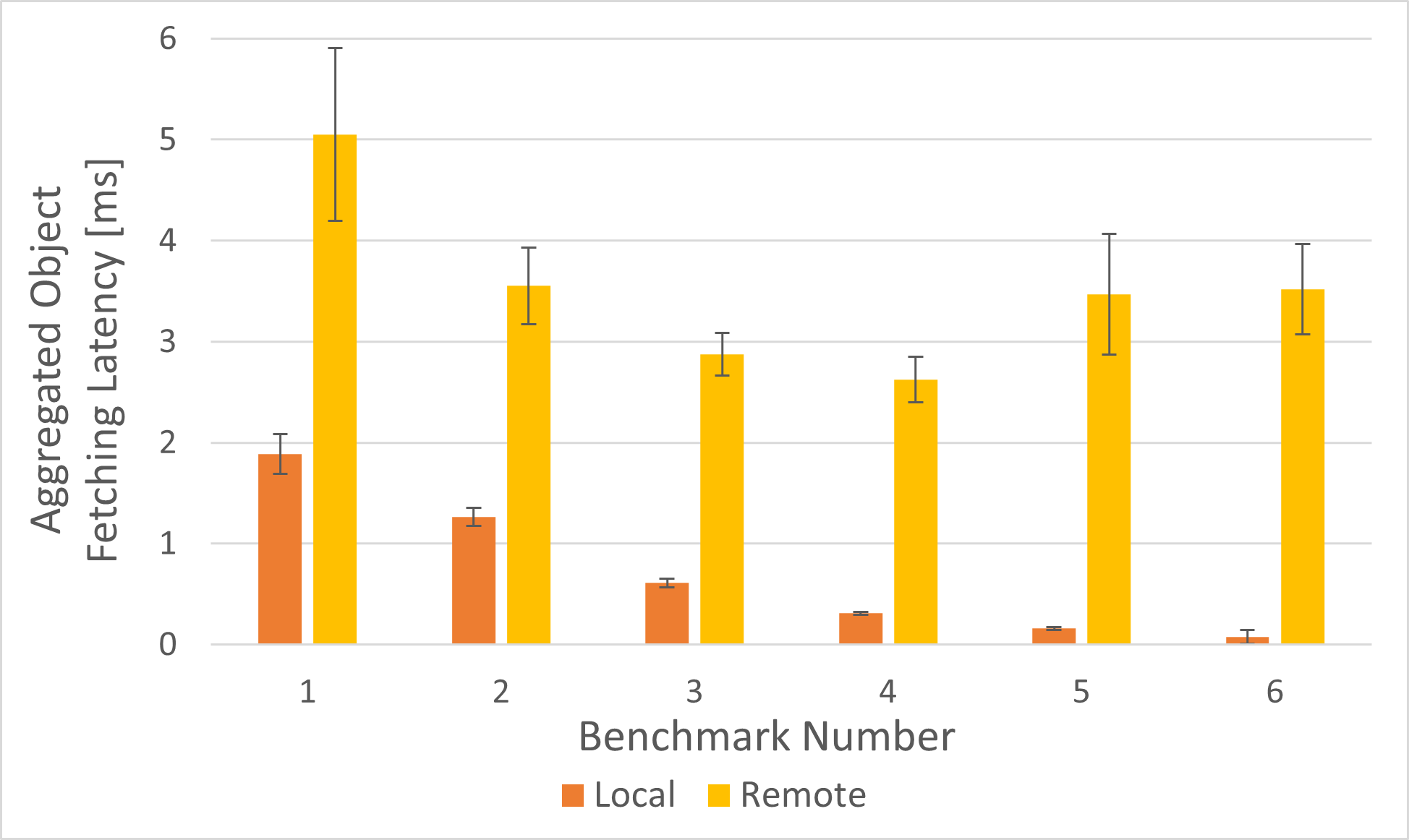}
    \caption{Plasma object buffer retrieval performance comparison}
    \label{fig:latency}
\end{figure}

Figure~\ref{fig:latency} plots the total object buffer retrieval latency per benchmark as measured from the time of the request to the reception of the last buffer. For local objects, the latency scales with the number of requested objects, ranging from 1.885 ms for 1000 objects to 0.075 ms for 10 objects. Remote objects incur a larger latency penalty due to the gRPC communication, ranging from 5.049 ms for 1000 objects to 2.624 ms for 100 objects. The complexity for the latter does also scale with the number of requested objects but the total latency is likely dominated by gRPC and its inherent network jitter, hence this is not clearly represented in the figure.

\begin{figure}[ht]
    \centering
    \includegraphics[width=1\linewidth]{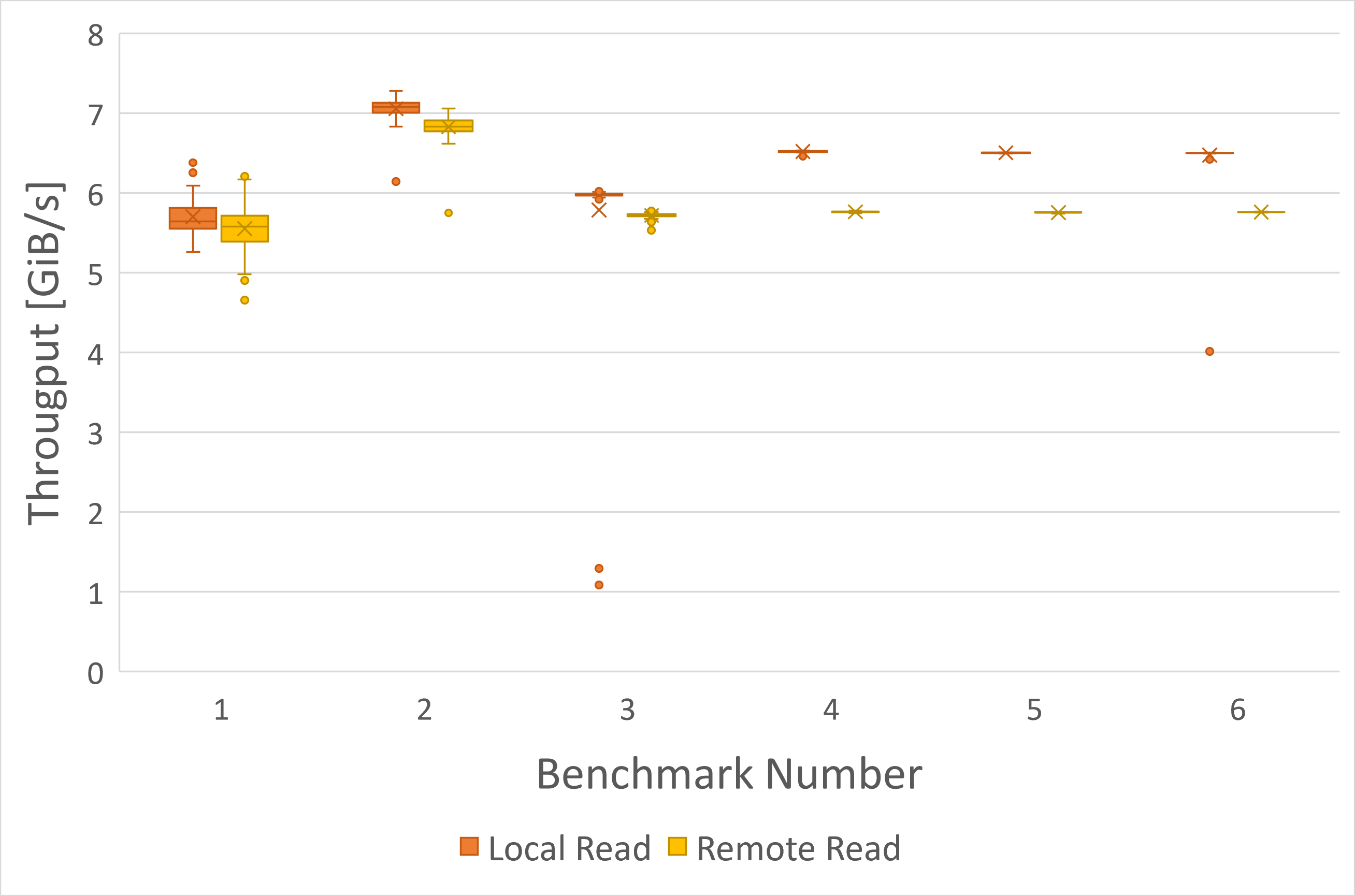}
    \caption{Plasma object buffer reading performance comparison (benchmark numbering in accordance with Table~\ref{tab:benchmarks})}
    \label{fig:bandwidth}
\end{figure}

Figure~\ref{fig:bandwidth} plots the throughput distribution per benchmark of consecutively reading the data from the requested buffers, including access latency. The results stabilize at 6.5 GiB/s for local objects and 5.75 GiB/s for remote objects in benchmarks 4-6. Benchmarks 1-3 display more variation (ranging from 5.5 to 7.1 GiB/s), which might indicate that the smaller objects in these benchmarks do not saturate bandwidth. Since the benchmarks run single-threadedly, they do not saturate full local and remote bandwidth completely anyway~\cite{thymesis}, however, the results are still competitive with the throughput of comparable state-of-the-art technology such as switched InfiniBand RDMA~\cite{rdma}.

As such, the results demonstrate that the proposed system can provide reading throughput of remote objects close to local performance (approximately 11.5\% difference). The fact that the penalty of remote disaggregated memory access is similar to comparable RDMA technology, means that the system delivers promising performance for big data applications. Object retrieval incurs a latency penalty in the order of milliseconds and marks a notable opportunity for improvement.

\subsection{Future work}
Although the microbenchmarks show promising results, the value of the system can be further demonstrated by extending our benchmarks to measure the system's performance in fully integrated solutions. Especially wide-dependency operations (commonly used in big data applications) pose an interesting subset for performance evaluation due to the ability of several nodes to operate on the distributed data in parallel.

As already outlined in Section~\ref{methodology}, several additional functionalities could improve the proposed framework. Firstly, a feedback mechanism for remote stores to exchange information about e.g. object usage will improve reliable operation. As previously discussed, tracking which objects are in use by clients shapes the object eviction policy, but this is not currently maintained across remote Plasma clients. A modified kernel module and disaggregated memory solution for exchanging remote object information could achieve this. Alternatively, additional RPC functionality could be added to attain the same functionality.

Moreover, in addition to the current RPC solution, which performs a remote call on every client request for unknown object identifiers, a caching mechanism for previously requested remote objects could be implemented. This would increase the performance of repeated requests for identifiers, which is dependent on system usage. This caching would require caution with tracking object usage by remote clients for the eviction policy and could result in corrupted object buffers if not handled carefully.

Furthermore, the currently presented system is implemented to accommodate a 2 node system. For rack-scale solutions, this needs to be modified to accommodate multiple nodes. The current system design allows for this modification.

In addition, there are several opportunities to further improve performance. Firstly, the simple replacement memory allocator was sufficient for demonstrative purposes but improved allocators generally have substantial impact~\cite{mallocs}. Additionally, the performance of remote object sharing could potentially be improved with an elaborate solution leveraging shared data structures in disaggregated memory. This allows direct look-up of remote objects in disaggregated memory and would likely improve performance but requires additional work.

Lastly, the proposed Plasma system was designed for ThymesisFlow, however, it is expected that novel future memory disaggregation systems such as IBM's Memory Inception~\cite{power10} could carry significant performance improvements~\cite{thymesis}. The modular design of the proposed system ensures that integration with future memory disaggregation technologies is possible.

\section{Conclusion}
\label{conclusion}
As the topic of memory disaggregation in data centers continues to gain importance, the demand for software frameworks that leverage the technology and quantify its potential increases. The current work set out to propose and demonstrate a novel type of in-memory object store based on the Apache Arrow Plasma API, which leverages memory disaggregation with the ThymesisFlow framework. The proposed framework pioneers using memory disaggregation in big data analytics by introducing novel programming models for handling very large data volumes in memory and leveraging distributed computing in an efficient and application-transparent manner.

The results demonstrate that the proposed system can deliver competitive single-thread throughput for remote disaggregated memory access with respect to local memory access ($\sim$5.75 GiB/s vs $\sim$6.5) and state-of-the-art RDMA technology. As such, the potential of the memory disaggregated object store framework is emphasized. Nevertheless, in its current state, there is a tremendous opportunity for improvements and future research in different forms. For example, the introduction of closely integrated memory disaggregation technology such as Memory Inception in IBM POWER10 and further integration into server- and rack-scale system designs. Further potential improvements for the proposed Plasma framework and its performance in fully integrated applications have been outlined for future investigation.

\section*{Acknowledgment}
We thank Felix Eberhardt and Andreas Grapentin (Operating Systems and Middleware Group of the Hasso Plattner Institute) for assisting in conducting our experiments and for providing comments on our work.

\bibliographystyle{unsrt}
\bibliography{references.bib}

\end{document}